# Pipeline Leak Detection Systems and Data Fusion: A Survey

**Uthman Baroudi, Anas Al-Roubaiey , and Abdullah Devendiran**
Computer Engineering Department, King Fahd University of Petroleum and Minerals, Dhahran, 31261 KSA

Corresponding author: Anas Al-Roubaiey (e-mail: anas@ kfupm.edu.sa).

The authors would like to acknowledge the support provided by King Abdulaziz City for Science and Technology (KACST) through King Fahd University of Petroleum & Minerals (KFUPM) and for funding this project No. 12-ELE2381-04 as part of the National Science, Technology and Innovation Plan.

**ABSTRACT** The pipeline leakage problem is a very challenging and critical issue. Solving this problem will save the nation a lot of money, resources and more importantly, it will save the environment. This paper discusses the state-of-the-art of leak detection systems (LDSs) and data fusion approaches that are applicable to pipeline monitoring. A comparison of LDSs is performed based on well-defined criteria. We have classified and critically reviewed these techniques. A thorough analysis and comparison of all the recent works have been provided.

## I. INTRODUCTION

Pipeline links are vital for a nation's infrastructure and social and economic well-being. Damaged water pipes deteriorate the quality of the transported commodity, resulting in taste, odor, and aesthetic problems in the water supply as well as public health problems [7]. Oil spills are known to destroy ecosystems and kill scores of aquatic organisms. Pipe damage results in other losses as well, such as increased operational and maintenance costs, loss of transported commodities (including oil, water, and gas), damage to property, disruption of service, disruption of industrial processes, increased environmental hazards, and imbalances in ecosystems. There is no doubt that efficient leak detection in pipelines can conserve a large amount of resources, save money, reduce carbon footprints, and achieve high levels of operational efficiency [28] [4].

Pipeline deterioration is caused by static factors, such as soil type, pipe material, size, etc., and dynamic factors, such as changes in pressure zones and climate. Little is known about the breaking modes of buried pipes, and the physical mechanism is not completely understood. The broad aspects of pipeline leak detection encompass physical modeling of the pipe in the soil, understanding the nature of pipe failure, empirical and/or statistical modeling of historical failures, inspecting pipes to identify stress factors, rating the pipe conditions, and modeling the deterioration to forecast future failures and residual life.

The length and size of the pipeline, type of product carried, proximity of the pipeline to a high consequence area, swiftness of leak detection, location of nearest response personnel, leak history, and risk assessment results, etc., determine the efficiency of a leak detection system. The parameters for the evaluation of a leak detection system (LDS) are derived based on API1995b and the Alaska Department of Environmental Conservation 1999. Generally, for any good LDS, the most important four criteria are [24], [38]: Reliability, Sensitivity, Accuracy, and Robustness, and these criteria are what we use in this work.

The following are some of the Characteristics of Leak Detection Systems under different environments, which were taken in consideration during this study:

*Type of fluids*: Pipelines transport a variety of fluids, such as gases, crude oil, petroleum products, steam, carbon dioxide, water, wastewater, etc.

*Type of operation*: Pipelines may operate in single-batch or multi-batch mode. In the *single-batch* mode of operation, pipelines operate continuously around the clock. In *multi-batch* mode, the pipelines function is based on a time schedule.

*Characteristics of leaks*: Leaks can occur suddenly or gradually depending on the causes and circumstances. *Sudden leaks* occur due to external damage, resulting in a significant change in the temperature, flow, pressure, etc. Gradual leaks may occur due to corrosion. Sudden leaks may be successfully detected using an internally based LDS. In contrast, gradual leaks have very low magnitudes, and dedicated equipment, such as externally based LDSs, may be required to identify such leaks.

***Operational phase***: Pipeline conditions vary. The *pumping* condition involves the transport of fluid, whereas in the *paused flow* condition, the fluid flow is zero. Sometimes, valves will be used to block the fluid flow in a given segment. This special flow phase is known as the *shut-in* or *blocked-line* condition.

The rest of the paper is organized as follows. In the next section, presents the leak detection techniques as mentioned in recent works. This section also includes discussion about WSN-based techniques that are used in monitoring pipelines. The third section discusses in detail the data fusion in pipeline monitoring. Finally, we conclude our work with recommendations and future directions in this subject.

## II. STATE OF THE ART OF LEAK DETECTION TECHNOLOGIES

The related work on leak detection systems as in [53] and [20] classified leak detection systems into visual, internal, and external based on the physical principles involved in the leak detection process. Monitoring can be continuous or non-continuous. In the classification by [34], non-continuous inspection includes acoustic and non-acoustic methods, whereas continuous monitoring includes measurement and model-based methods. Reference [49] classified technologies based on the area of inspection, such as internal pipe surface, pipe wall integrity, and pipe bedding/void conditions. References [35], [44], and [51] classified leak detection systems into non-technical and hardware- and software-based methods. Non-technical methods do not involve any devices and use only natural senses, such as hearing and smelling, to identify a leak, whereas the technical methods use special devices to identify leaks; in the hardware methods, these devices include liquid sensing cables, vapor sampling, etc., and in the software methods, these devices include negative pressure waves, pressure point analysis, etc. Reference [3] divided the leak detection systems into visual, physical, acoustic, ultra-spectrum, and electromagnetic. A similar classification by [22] divides LDSs into visual, acoustic, and Electromagnetic-Radio Frequency (EM-RF) techniques. Fig. 1 depicts the LDSs classification. Recent high level abstraction classification for water distributed network leak detection in [56], which classify LDS into transient, model, and data based approaches.

LDSs can be broadly classified into continuous and non-continuous monitoring systems. In non-continuous monitoring systems, the inspection is performed at regular intervals. Depending on the mode of inspection, pipeline operations can either continue or need to stop. For example, visual inspection or a helicopter survey does not require pipeline operations to be stopped, whereas an intelligent pigging system may require the operations to be stopped. The remote sensing of liquid hydrocarbons using aircraft mounted gas remote sensing is given in [48]. This system detects evaporative plumes from pools of oil, gasoline, condensate, or pentane. Continuous monitoring systems monitor pipelines around the clock and are based on a physical principle. This approach can further be classified into external and internal systems.

### A. VISUAL TECHNOLOGIES
***Visual Manual Inspection***: Visual inspection requires the manual patrolling of the pipeline for leaks. Patrolling can be performed by any means (e.g., walking, in a vehicle, or from a helicopter). The operator examines the area for stains or other evidence of leaks. The leak detection capability depends on the ability of the inspection team, frequency of inspection, and the size of the leak. Limited for reachable pipelines; and not real-time detection, which has a negative effect in terms of loss of oil and gas as well as environmental pollution.

***Smoke/gas testing:*** A smoke bomb is placed inside a water pipe with a blower to push the smoke. The smoke filters out through any cracks, thus exposing them. Water utilities used *Formier10* gas (10% hydrogen and 90% nitrogen) for approximately 20 years [18]. Hydrogen is a very lightweight gas and easily escapes through small cracks. The time taken for the gas to reach the surface depends on the depth of the pipe, soil conditions, and size of the leak. The gas detector is sensitive to small leaks. This method is not usable in large pipe mains due to the larger volume of gas required.

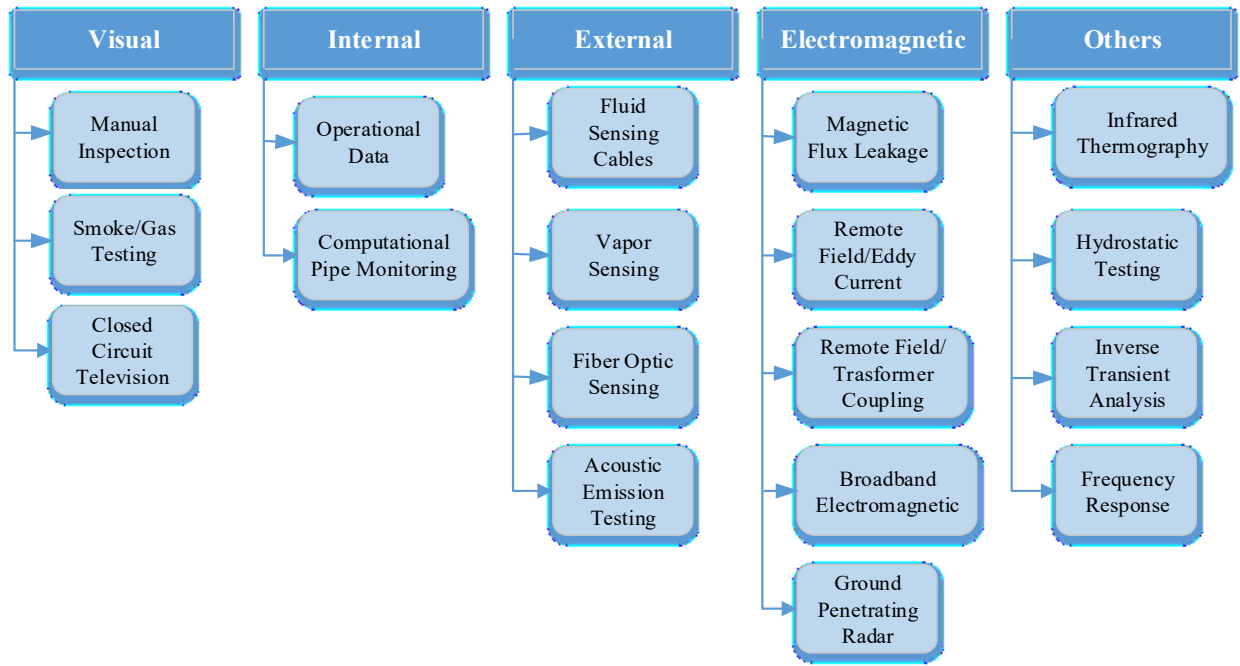

**FIGURE 1.** Classification of LDSs.

***Closed Circuit Television (CCTV)*:** CCTV technology typically involves the use of a video camera, lighting source *and* event recording software. The camera is passed through the pipe and records the interior surface. The operator later looks for defects in the pipe from the recorded images.

*B. INTERNAL SYSTEMS*

Internal systems use field sensors to monitor the operational and hydraulic conditions of the pipeline, e.g., measurements of the flow, pressure and temperature. The normal working parameters of the pipeline are determined either manually by pipeline controllers or based on sophisticated algorithms and hydraulic models, e.g. in [59], which function with particle swarm optimization AI technique to get accurate detection and localization.

A difference between the measured and predicted operational parameters indicates a leak. Typically, the remote field sensors provide data to a centralized monitoring station, where the data undergoes filtering, signal processing and modules with leak detection algorithms to identify a leak. Internal systems generally do not require the installation of extensive hardware throughout the pipeline. Fig. 2 illustrates different types of internal LDS techniques.

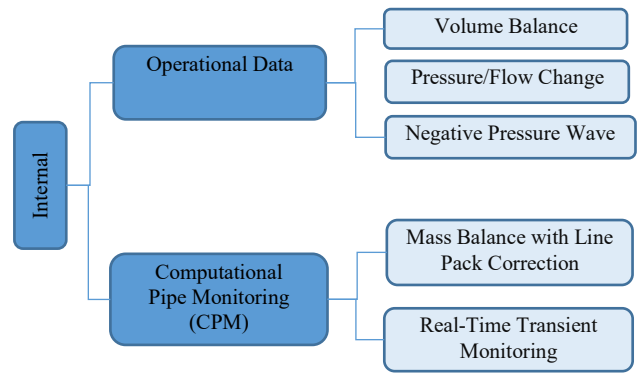

**FIGURE 2.** Internal LDSs

### 1) REGULAR OR PERIODIC MONITORING OF OPERATIONAL DATA

***Volume balance:*** Volume balance identifies the imbalance between incoming and outgoing volumes. Volume balance can detect catastrophic failures; however, its usage is rare due to its limited performance.

***Rate of pressure/flow change:*** The rate of pressure or flow change is based on the principle that a leak causes a rapid change in pressure. First, a sudden pressure drop can also be due to transient conditions. Filtering techniques need to be used to differentiate operational conditions from leak conditions. Second, pressure waves damp out as they traverse a longer length and thus additional pressure sensors need to be installed along the pipelines. This method is only effective for large leaks, and transient conditions may trigger false alarms.

*Negative Pressure Wave (NPW):* Sudden leaks create a negative pressure wave or rarefaction wave, which prop- agates in both directions from the leak. NPW is easy to install and maintain and capable of continuously monitoring pipelines. However, the system cannot distinguish between leak scenarios and normal operations, thereby giving raise to false alarms. ATMOS, a novel technique that was developed recently and is based on the rarefaction method, shows tolerance to transient, shut-in, and slack flow conditions, thus triggering few false alarms [13].

2) COMPUTATIONAL PIPELINE MONITORING (CPM)
CPM detects hydraulic anomalies in pipeline operating parameters [5].

*Mass balance with line pack correction:* The changes to a line pack are observed by various sensors, e.g., pressure, temperature and densitometers, at multiple locations between the inlet and outlet flow meters. The pipeline is divided into multiple segments based on certain factors, such as elevation profile, location of instruments, desired level of accuracy, etc. The changes measured by various sensors are adjusted in the mass balance to account for transient flows, anticipated fluid changes, and other flow conditions. The capability depends on the selection of the alarm set points, repeatability of the instrumentation, skill of the pipeline controller, etc. The method is retrofit table but less adaptable to complex pipeline configurations.

*Real Time Transient Modeling (RTTM):* The parameters derived from a simulation model are compared with actual field data to look for discrepancies. Leaks occurring under all flow conditions can be modeled using this software, and small leaks can be detected in seconds. However, RTTM needs extensive training and skilled workers to operate and maintain.

### C. EXTERNAL SYSTEMS
External systems use local sensors to detect fluids escaping from pipes. Impedance methods use cables with fiber optic or electro-chemical detection to sense liquids. Sniffing methods depend on vapor sensing through tubes. Acoustic methods depend on sensing noises induced by leaks. These systems are highly sensitive to leaks and can accurately locate them [24], [21]. However, due to the high costs, these methods are employed only in sensitive locations or for short pipeline segments [41] [38].

*Liquid sensing cables:* Similar to optical fiber methods, *liquid sensing probes* or cables are laid throughout the pipeline. Leaking fluids come into contact with cables and change their electrical properties, such as impedance, electrical resistance, dielectric constant, etc. A dedicated evaluation unit connected to the cable identifies the changes to the cable and detects a leak. Liquid cables can continuously monitor and accurately locate leaks. As with optical fibers, cable replacement may be required after a leak occurs.

*Vapor sensing cables* [21]: In the vapor sensing method, a highly permeable, pressure-tight air tube is fitted along the entire length of the pipe. When a leak occurs, the leaked material diffuses into the tube due to the concentration gradient. After a certain time, an accurate image of the substance surrounding the tube is obtained. A column of air that is pumped at constant speed passes through a gas sensor, and the substance produces a peak, indicating a leak. The increase in the gas concentration produces a leak peak, and the height of the peak is proportional to the concentration of the substance, which is an indicator for the leak size (Fig. 3). Initially, electrolytic cell is used to inject a test gas from the end of the detected line to pass through the entire length of the pipe. The detector unit marks the start peak and the end peak to calculate the length of the pipe. When a leak occurs, the ratio of leaked distance to the overall distance is calculated to identify the location of the leak.

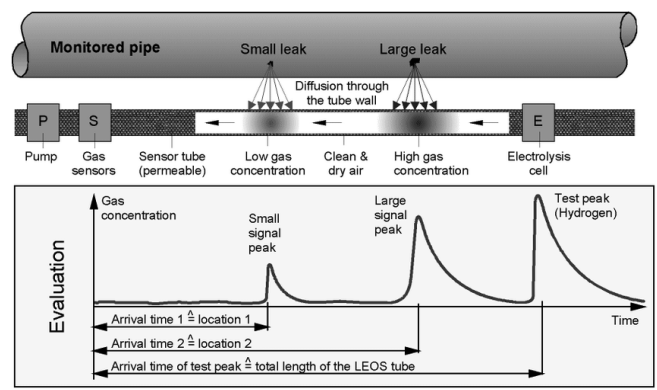

**FIGURE 3.** Vapor sensing tube [21]

*Fiber optic sensing cables:* In this method, a fiber op- tic cable is installed along the entire length of the pipeline. When a liquid comes into contact with the cable, the transmission characteristics of the fiber change.

While a pulsed laser propagates through the fiber, any changes to the density or composition of the fiber cause the light to scatter backwards. Spectral analysis reveals the temperature profile, leading to leak detection and localization [21], [33]. The process is depicted in Fig. 4. Cable replacement may be required after a leak. Recent advances in fiber optic sensors include quasi-distributed sensing, e.g., integrated Bragg gratings, and distributed sensing, e.g., interferometry and fiber sheath sensors ([52], [25]).

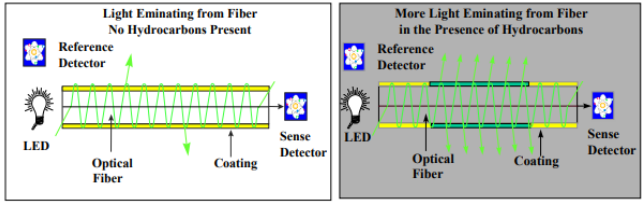

**FIGURE 4.** Leak detection using a fiber optic cable [33].

When a pipeline leak occurs, the liquid moves from a high-pressure area to a low-pressure area and a turbulent flow is generated. This flow generates a characteristic sound that can be picked up by a specially designed hydrophone. Using sophisticated software, LeakACO detects this signal, analyzes it, and evaluates the measurement results, thus identifying and providing the location of the leak.

*Acoustic Emission (AE) testing*: When a pipeline leak occurs, turbulent liquid flow occurs in a high-pressure to a low-pressure area, creating a low frequency sound signal. Acoustic sensors affixed to the outside of the pipe, e.g., accelerometers, hydrophones, piezoelectric transducers, etc., pick up these signals. The deviation of the sound signals from a baseline fingerprint triggers a leak alarm. The received signal is stronger near the leak site, enabling localization of the leak. For pipes, such as PCCP, wire breaks release energy and cause a series of discrete events. AE monitoring of wire breaks is limited only to on-going wire events and cannot be used to detect already broken wires. Another recent technique [58] using a combination of the ultrasonic sound and flow rate signals in order to detect and localize a small pipeline leak.

### D. ACOUSTIC TECHNOLOGIES
*Listening rods:* Acoustic signals from leaks propagate through the soil to the surface. Listening rods placed on the ground in the proximity of the leak pick up these noises and determine the location of the leak.

*Inline acoustic leak detection*: Inline acoustic leak detection sensors pass through pipes while in service and detect sounds due to leaks. The equipment can be tethered or free swimming. This method detects leaking joints and welds very well.

*Leak noise correlator*: Acoustic sensors placed on either side of a suspected leak transmit leak signals to a noise correlator. The correlator is typically a computer that analyzes the input sound spectrum and pinpoints leaks based on the time lag and sensor-to-sensor spacing.

*Sonar or ultrasound*: A major drawback of ultrasonic devices is that they cannot be operated above and below the water line simultaneously. To overcome this limitation, CCTV and sonar can be mounted on the same carrier vehicle so that CCTV can capture the information from above the waterline, while sonar captures the information below the waterline to account for the shortcomings of both systems [50].

### E. ELECTROMAGNETIC SYSTEMS
*Magnetic Flux Leakage (MFL):* MFL analyzes the flux leakage in a magnetic field when magnetized by strong, powerful magnets. A flawless pipe exhibits a homogeneous magnetic flux distribution, while a damaged pipe causes a flux leakage, as shown in Fig. 5. The detection system also consist of a smart tool that can reflect the changes in the flux distribution in case of leakage or corrosion; this tool acquire the measurements from a sensor, which is placed between the poles of the magnet. DC inspection of pipes can be performed using Hall Effect devices and magneto resistive materials, while AC inspection can be performed using pick up coils. This testing mode is non-invasive and accurately detects cracks, corrosion, and the thinning of pipe walls. However, MFL is usable only on ferrous pipes and requires access to the surface of the pipe. The analysis of test results requires experienced personnel. Traditional MFL only detects defects perpendicular to the magnetic field and cannot identify defects parallel to it. To overcome this, a new inspection method called Traverse Field Inspection (TFI) is employed in the Spiral MFL tool [27].

*Remote Field Eddy Current (RFEC)*: In BEM, a solenoid exciter probe generates pulsed eddy currents and magnetic flux lines within the pipe. Anomalies such as cracks or defects disrupt the current flow, which is captured by a receiving probe placed at a distance of 2.5 pipe diameters. The contour maps obtained after intensive post processing reveal the corrosion and thickness of the pipe wall [50].

*Remote Field Transformer Coupling (RFTC):* RFTC detects any broken wires in pre-stressed concrete cylinder pipes (PCCP) and holes or perforations in the steel used in PCCP [50].

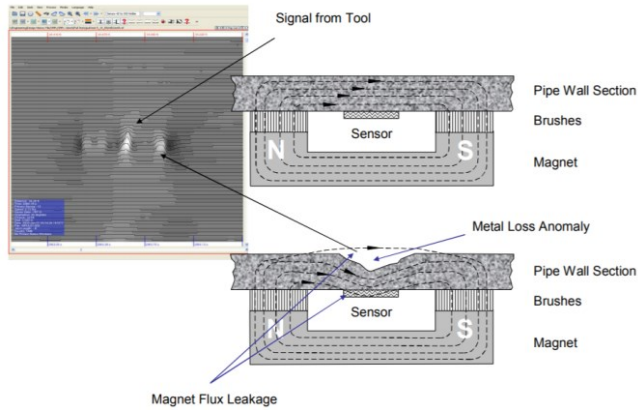

FIGURE 5. Principle of Magnetic Flux Leakage [27].

***Broadband Electromagnetic (BEM):*** In BEM, a primary winding or exciter coil generates a short burst of pulsed waves in the broadband frequency range. Eddy currents are induced in the adjacent ferrous conductive material shortly after the excitation pulses have been turned off; these eddy currents create a time varying magnetic field. The varying magnetic field induces a time varying voltage on the secondary winding or the receiver coil, which is correlated to the thickness of the pipe. BEM is similar to RFEC, but the signal transmitted covers a broad frequency spectrum [50]. BEM is immune to electromagnetic interference and differs from other electromagnetic inspection methods because of its frequency independence.

***Ground Penetrating Radar (GPR):*** GPR transducers radiate a short burst of varying radio frequencies into the ground and identify buried objects based on the scattering of the EM waves. The propagation of EM waves in soils is governed by parameters such as permittivity, magnetic permeability and conductivity. The occurrences of leaks increase the moisture content of the soil nearby and cause dielectric variation. Reflections occur at the interfaces between media with different electrical properties. The time lag between the transmitted and reflected waves determines the depth of the objects. An array of antennae attached to a survey vehicle driven along the transmission main detects the pipe anomalies. A three-dimensional (3D) GPR image is obtained using the raw field data after software processing. Example GPR data before and after interpretation are shown in Fig. 6. Highly skilled expertise is needed to interpret the data. From the perspective of system design, GPR falls into three main categories ([32]):

1. Time domain: Impulse GPR
2. Frequency domain: frequency modulated continuous waveform (FMCW), stepped frequency continuous waveform (SFCW), and noise-modulated continuous waveform (NMCW) GPR
3. Spatial domain: Single frequency GPR

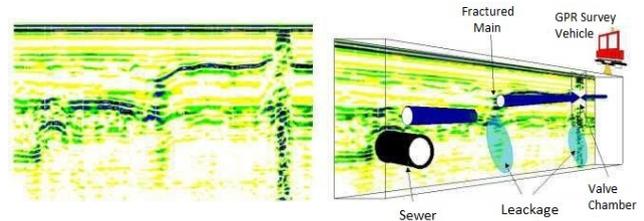

FIGURE 6. GPR data before and after interpretation [32].

### F. PIPELINE MONITORING USING WSN

A sensor node in Wireless Sensor Network (WSN) typically consists of transducers (to determine variations in temperature, pressure, strain, etc.), analog-to-digital converters, signal processing, power sources, memory, etc. Typical sensors used in pipeline monitoring are discussed in this section.

***Acoustic sensors:*** Acoustic sensors are based on the principle that a liquid escaping through a hole in a pipeline produces a detectable sound. Acoustic sensors are easy to install and maintain and can continuously monitor a long pipeline. An important drawback is their high susceptibility to noise sources, such as system noises, environment noises, radio chatter, wind, Doppler effects, etc. To eliminate system noises, various techniques, such as band pass filtering [17], Fast Fourier Transform (FFT) and time-averaging Wigner-Ville distribution [54], can be used. Acoustic sensors can be used along with other sensors to overcome these limitations. In [46] they used piezoelectric sensors along with acoustic sensorsto identify leaks and other pipe defects. In [60], they proposed to extract time-domain statistical features from the acoustic sensors instead of the amplitude and the frequency domain related features.

***Piezoelectric sensors:*** Piezoelectric sensors (PZT) or lead zirconate titanate sensors can monitor the physical properties of pipelines, such as pressure, acceleration, vibration, acoustic waves, etc., and convert them into electric signals. The strength of the signal is determined by factors, such as the amount of energy released, distance from the monitored event, orientation of the sensor, transmission media, etc. [40], [39], [29]. Piezoelectric sensors are a suitable candidate for pipeline monitoring because they exhibit high tolerance to harsh conditions and are not sensitive to electromagnetic radiation. However, piezoelectric sensors are not free from generating false alarms because a sensor deployed to detect one physical quantity may be affected by another, e.g., a pressure sensor may be affected by vibrations in the pipeline. To compensate for this limitation, secondary sensors can be used where the pressure sensors are used along with accelerometer piezoelectric sensors to detect pressure transients.

*Chemical sensors:* Chemical sensors determine a defect based on a change in chemical composition. Oxygen, carbon monoxide, and mercury vapor sensors are some examples of chemical sensors. The parameter to be detected and the resultant effect vary between different types of chemical sensors. For example, mercury sensors cause a change in resistance in the case of a gas leak [10]. In another approach, the weight of the material changes considerably. Chemical sensors are very handy device in hazardous environments. Reference [57] demonstrate recent advances in using WSNs in oil and gas industry, and provide new directions in this subject.

WSNs provide effective solutions for pipeline monitoring, due to its low cost, flexibility and ease of deployment in inaccessible terrain. However, some design issue need to be addressed before selecting wireless deployment. The major design issues that should be taken in consideration when using WSN for monitoring pipelines are: power source, communication standard, node antenna, communication protocol, localizations, network reliability, density of sensor nodes, packet loss control and network congestion control.

## III. DATA FUSION IN PIPELINE MONITORING

This section classifies and describes the data fusion approaches in pipeline monitoring, and provides the relevant architecture models.

### A. CLASSIFICATION AND ARCHITECTURE MODELS

The fusion of data from multiple sensors, called multi-sensor data fusion, provides more information than a single sensor. Multi-sensor data fusion can also include fusing overlapping measurements from the same sensor obtained at different times. Data fusion improves performance in at least four ways: representation, accuracy, certainty, and completeness [1]. Durrant-Whyte classified data fusion based on the relationship among the sources, such as complementary, competitive, and co-operative [15].

*Complementary:* Non-redundant data from different sensors can be fused to provide a complete view.
*Redundant (competitive):* The same pieces of data from a single sensor or multiple sensors can be fused to increase the associated confidence.
*Co-operative:* Different data can be fused to provide a realistic view.

The abstraction levels of the input and output in the fusion process, including the measurement, signal, feature, and decision, can also form a basis for classification. Reference [31] applied these levels to classify fusion into signal fusion, pixel fusion, feature-level fusion, and symbol fusion. Boudjemaa and Forbes classified data fusion based on time, domains, attributes, and sensors [7]. DaSarathy classified data fusion according to its input and output characteristics [12]: DaI-DaO (Data Input/Data Output), DaI-FeO (Data Input/Feature Output), FeI-FeO (Feature Input/Feature Output), FeI-DeO (Feature Input/Decision Output), DeI-DeO (Decision Input/Decision Output).

Data fusion architecture models can be *data based*, such as JDL [36] and DaSarathy [11], *activity based*, such as Boyd control loop, intelligence cycle, and the omnibus model [6], or *role based*, such as object-oriented and Frankel-Bedworth [19].

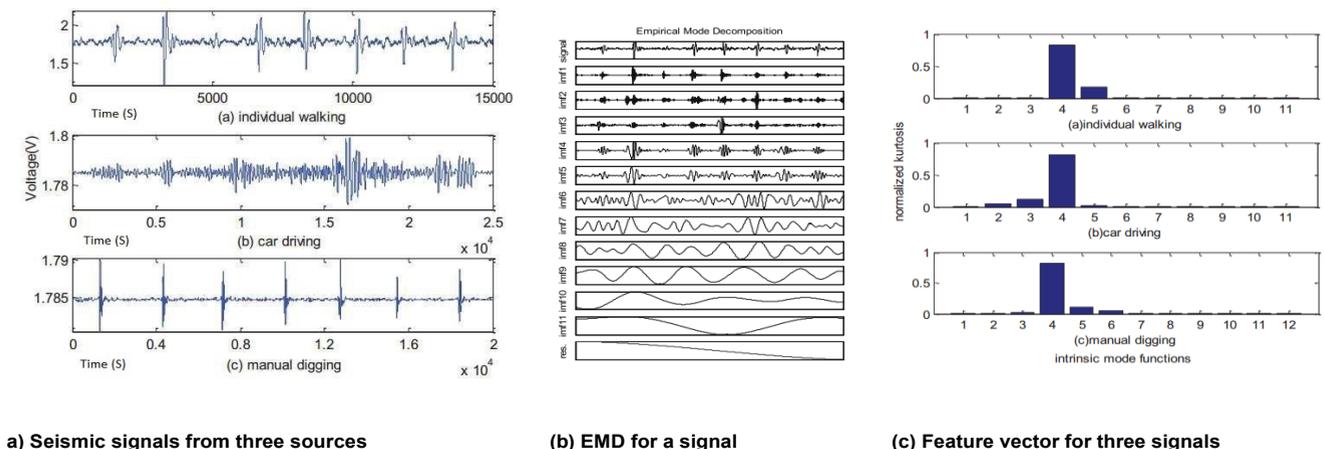

(a) Seismic signals from three sources     (b) EMD for a signal     (c) Feature vector for three signals

**FIGURE 7.** Signal Processing using EMD, HHT and Kurtosis [47].

Reference [26] provided a data centric taxonomy of data fusion methodologies and discussed the fusion of imperfect data, the fusion of correlated data, and the fusion of inconsistent data. Imperfect data can be fused using probabilistic, evidential, fuzzy reasoning, possibility theory, rough set theory, random set, and hybridization approaches. The fusion of correlated data can be achieved using correlation elimination and correlation presence. The fusion of inconsistent data focuses on removing the outliers, disorders, and conflicts.

### B. EXISTING DATA FUSION APPROACHES

Some of the Data Fusion (DF) schemes applied in pipeline monitoring are discussed as follows:

***Homogeneous DF of seismic pulses***: Reference [47] used Dempster-Shafer (D-S) method to fuse data from multiple seismic sensors in a proactive pipeline monitoring system. The data fusion increased the accuracy of the decisions by 8-25%. To detect the seismic pulses, geophones were deployed along the length of the pipeline at a depth of approximately half a meter, with varied sensor spacing. Different sources, such as people walking, driving a car, manual digging, etc., generated seismic signals with different frequencies, as shown in Fig. 7a. The signals were then amplified, filtered, and A/D converted, followed by the extraction of the features. Fig. 7b shows the decomposition of the original time series data into intrinsic oscillation mode functions (IMF) using empirical mode decomposition (EMD). Each IMF component was subjected to a Hilbert-Huang Transform (HHT) to obtain the amplitude and frequency. Normalized Kurtosis gives the feature vectors of different signals, as shown in Fig. 7c, and extracts the features of each target.

***Data driven framework using DF***: Reference [55] proposed a data driven framework that used piezoelectric wafers to generate and sense ultrasonic waves. Multiple signal processing techniques were applied to extract as many as 365 features. The wave patterns were then checked using an adaptive boosting algorithm and five machine learning classifiers for damage detection. The system was shown to exhibit an average accuracy of 84.2-89%. As shown in Fig. 8, the received signals were preprocessed to remove low/high frequency vibrations via a band pass filter of the range from 190 to 450 kHz. By normalizing the signals, the ambient effects were contained to a certain degree. Various signal processing methods, such as the wavelet transform, Hilbert transform, Mellin transform, etc., were then applied to extract features.

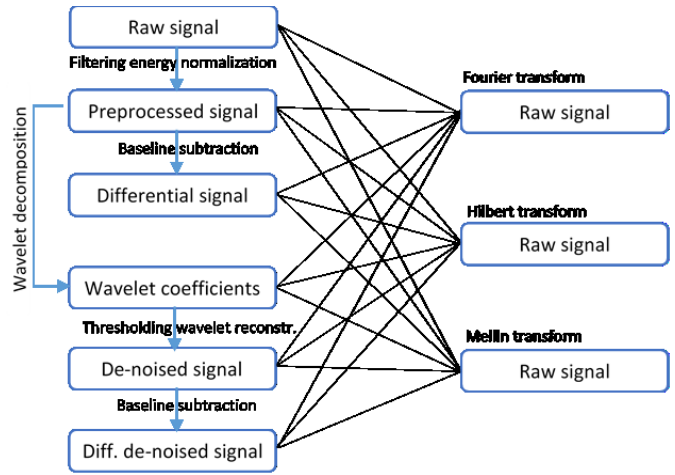

**FIGURE 8**: Signal processing to extract features [55].

***DF based on voting strategy***: Ultrasonic waves, despite their ability to monitor across long distances, suffer from sensitivity to environmental changes, such as wet conditions, surface vibrations and temperature extremes.

A voting strategy-based data fusion in a spatially distributed sensor network is given in [37]. Certain features, such as the normalized mean squared error (MSE), correlation coefficient, curve length, loss of local coherence, etc., were used to detect the damage. For data fusion, the independent decisions were fused to arrive at the outcome, i.e., decision-level data fusion was employed. The features for all monitored signals were compared against the threshold, as shown in Fig. 9. A lower false alarm rate means that the result is highly accurate. For each transducer pair, a voting strategy was used to increase the credibility. The system seemed to increase the detection probability to more than 90% and reduced the false alarms to under 5%.

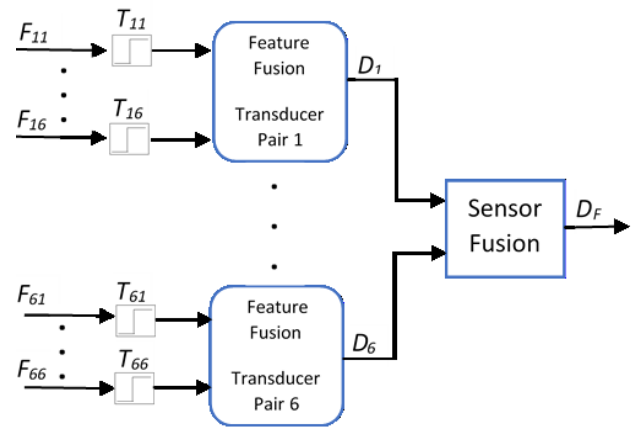

**FIGURE 9**: Sensor and Feature fusion at the decision level [37].

***Fusion of GPR and EMI for buried pipes***: A multisensory system was used to fuse data from seismic, GPR, and EMI sensors to reduce false alarms in landmine detection [45]. The responses of the three types of sensors to the soil condition differed from each other. EMI sensors were sensitive to soil conductivity, while the seismic sensor was sensitive to the difference between the mechanical properties of the soil and the landmine; GPR was sensitive to dielectric properties. A good use of the complementary features from sensors reduced the false alarm rate significantly. Reference [2] proposed a multisensory data fusion architecture to assess the locations and structural conditions of the buried pipes.

Using a combination of ground penetrating radar and electromagnetic waves, inspection can be performed without draining the pipelines. GPR can detect the presence and depth information of buried pipes but cannot distinguish metallic and plastic pipes. Electromagnetic sensors can detect the condition of metallic pipes but cannot measure the depth. A data fusion algorithm that is based on artificial neural networks and uses a combination of inputs from GPR and EMI can detect and classify various defects, such as major cracks and leaks in pipelines. However, this architecture is only conceptual, and the implementation results are not yet known.

**Heterogeneous DF of NDE methods using geometric transformation:** In [47] they employed a neural network-based geometric transformation algorithm to fuse data from images obtained from three NDE methods: IR thermal imaging, magnetic flux leakage (MFL), and ultrasonic testing (UT). Given a training data set, the radial basis function identified redundant and complementary features using artificial neural networks (ANN). Redundancy increased the reliability of defect characterization by identifying the common information in different NDE methods. Complementarity improved the accuracy of defect characterization by identifying the defect characteristics unique to each inspection method.

Let $x_1(r, c_1)$ and $x_2(r, c_2)$ denote the two different NDE images, where $r$ represents the redundancy feature and $c_1$ and $c_2$ represent the complementary features. Then, the redundancy ($hr$) and complementary information ($gr$) are defined in equations 1 and 2:

$$f\{x_1(r, c_1), x_2(r, c_2)\} = h(r) \tag{1}$$

$$f\{x_1(r, c_1), x_2(r, c_2)\} = h(c_1, c_2) \tag{2}$$

The redundant relation between the data are given by equation 3.

$$h_1(r) \diamond g_1(x_1) = g_2(x_2) \tag{3}$$

In equation 3, $\diamond$ represents a homomorphic operator and $g_1(x_1)$ is a radial basis function that takes the training data set as an input and outputs the best function approximation for $x_1$. $g_2(x_2)$ is a conditioning function and application dependent. For example, if the data $x_2$ is spread over a wide range, a logarithmic function can be used for $g_2(x_2)$. From equation 4, if the homomorphic operator is chosen as an addition operator + and $g_2$ is assumed as the identity function, then $h_1(r)$ is given by equation 4.

$$h_1(r) = x_2 - g_1(x_1) \tag{4}$$

Similarly, the complementary relationship can also be defined as in equation 5 and $h_2(c_1, c_2)$ is obtained, but the neural network is trained with the complementary data.

$$h_2(c_1, c_2) \diamond g_1(x_1) = g_2(x_2) \tag{5}$$

The training of the artificial neural network with diverse and sufficient NDE signatures is essential for improved fusion. The data fusion for different NDE combinations is shown in Fig. 10.

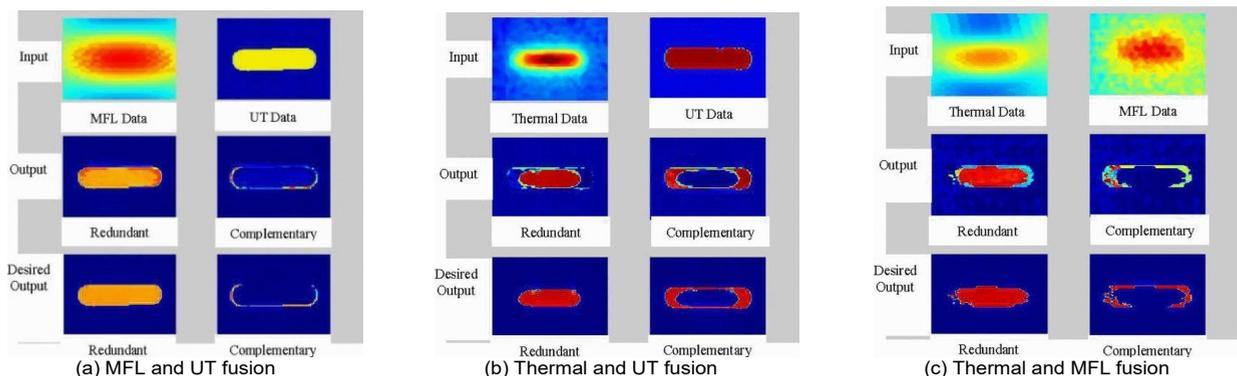

**FIGURE 10.** Data fusion combinations for MFL, UT and Thermal Imaging [47].

## IV. COMPARISON and ANALYSIS

As explained in the previous sections, the leak detection problem is a very complex and multidisciplinary problem. Hence, in order to conduct a fair comparison among available techniques, it requires identifying multiple criteria. Through our study, we compile a list of criteria to compare these techniques. Moreover, these criteria are classified into three categories: *Technical*, *Operational* and *Economical*.

Technical criteria are related to the technical performance of a specific technique in achieving these criteria such as leak size, response time, leak location estimate, false alarms, robustness. For example, "leak size" is an important feature which determines the sensitivity of a specific technique to detect the leak; is it able to detect a small leak? Another critical feature is the ability to localize the leak; some technique can detect the leak but it is unable to localize.

On the other hand, operational criteria are related to functioning features of a specific technique while it is on operation such as: shut-in condition, availability, complex configuration, simplicity, ease of testing and ease to maintain. For instance, shut-in condition feature tells whether a specific technique can work while the pipeline is on operation or off.

Finally, economical criteria are related to economical features of a specific technique while it is on operation such as ease of training and cost. Table 1 describes the criteria used in the comparison study.

Table 2&3 compare internal and external LDS system characteristics based on the features stated above, respectively. In addition, Table 4 summarizes and compares the capabilities and limitations of different existing techniques.

TABLE 1
DESCRIPTION OF COMPARISON CRITERIA

| Class | Criteria | Description |
|---|---|---|
| Technical Features | Response Time | The time required by the LDS to report the existence of a leak and issue an alarm. |
| | Released Volume Estimate | The ability to measure the released liquid during the leakage period |
| | False Alarms | The rate of inaccurately declaring the existence of a leak. |
| | Robustness | "the ability of the leak-detection system to function and provide useful information, particularly under changing conditions of pipeline operation or situations where data is lost or suspect" [62]. According to API 1155, it is defined as a measure of the LDS ability to continue to operate and provide useful information, even under changing conditions of pipeline operation, or in condition where data is lost or suspect [63]. |
| | Complexity | A measure of the complexity of the technique in terms of computation. |
| | Complexity Configuration | A measure of the configuration complexity. |
| | Noise susceptibility | The sensitivity of LDS to function properly in the existence of external noise |
| | Leak Sensitivity | A composite measure of the size of leak that a LDS is capable to detect, and the time required for the system to issue an alarm [64]. |
| | Affected by multiphase or multi-component | The sensitivity of LDS to multiple changes in pipeline operation conditions. |
| | Location Estimate | The ability to localize the leak location |
| | System Transients | The sensitivity to transient waves due to sudden changes such as turning a pump off when a leak occurs. |
| Operational Features | Shut-in Condition | "a production cap set lower than the available output (of an oil producing site)" [63]. |
| | Slack Condition | Slack flow occurs in a pipeline when the pipeline pressure falls below the vapor pressure of that liquid. This may cause in accuracy in leak detection [61]. |
| | Ease of Use | A measure of usability of a given LDS |
| | Cont. Monitoring | A measure of whether LDS can monitor continuously or on demand. |
| | Ease of testing | A measure of ease of handling the test using a given LDS. |
| Economical Features | Ease of Retrofit | A measure of LDS arability to work/cooperate with new techniques. |
| | Ease of training | A measure of ease of training operators to work on a given LDS. |
| | Maintenance Requirements | A measure of conducting/ preparing for regular maintenance. |
| | Cost | Typically, the external LDS cost much more than internal LDS. |

## IV. CONCLUSION

Each leak detection system is a unique system and designed based on the pipeline for which it is developed for. The choice of LDS should be based on a fit-for-purpose approach. The operating parameters, such as the pipeline size, length, instrumentation design, etc., dictate the applicability of an approach. The time taken to detect a leak, number of false alarms, accuracy of the installed instrumentation, and many other factors influence the performance of a leak detection system. The capabilities of each LDS and the degree to which they mitigate the risks discussed in this paper can be used as a guideline when choosing a leak detection approach. The rule of thumb is that field test results from similar applications always provide the best recommendations. The use of information from multiple LDSs increases the detection accuracy. There are multiple uncertainties in data sources, including hydraulic noise, errors in analog-to-digital conversion, the non-repeatability of field sensors, data communication errors, timing, drift, transient conditions, etc. The uncertainty in the data is a crucial issue because, without proper inputs, a correct output cannot be achieved no matter how efficient the filtering, signal processing or data fusion algorithms may be. The quality of a data fusion framework depends on good input data and the performance of the fusion system. A thorough review of the available historical data regarding pipe performance and failure can give greater insight into applying data fusion and accurately predicting pipe deterioration. Future directions for developing and improving leak detection systems are varied from increasing the accuracy, by minimizing the false alarms and precisely determining the leak position, to using the new technologies in improving these types of monitoring systems, such as machine learning, Internet of Things, and drones monitoring.

TABLE 2
FEATURES COMPARISON OF INTERNAL LDS METHODS

| Internally Leak Detection Methods | Technical Features* | | | | | | Operational Features | | | | | | | | Economical Features | | | |
|---|---|---|---|---|---|---|---|---|---|---|---|---|---|---|---|---|---|---|
| | Response Time | Released Volume Estimate | Existing Leak | Leak Sensitivity | Location Estimate | System Transients | Shut-in Condition | Slack Condition | False Alarms | Robustness | Cont. Monitoring | Complex Configuration | Complexity | Ease of testing | Ease of Retrofit | Ease of training | Maintenance Requirements | Cost |
| Volume Balance | Minutes to hours | no | yes | 1-5% | no | no tolerance | no | no | frequent | average | part time | no | simple | easy | easy | easy | Low | average |
| Rate of Pressure /Flow Change | minutes | yes | no | 5% | yes for large leaks only | some tolerance | yes | no | frequent | low | part time | no | complex | more difficult | easy | difficult | Low | higher |
| Volume Balance /Line Pack Compensation Using Actual Pressure measurements | minutes | no | yes | 1% | no | better tolerance | yes | no | less frequent | average | yes | no | less simple | easy | not easy | easy | Med | higher |
| Volume Balance /Line Pack Compensation Using Dynamic Computational Model | minutes | no | yes | 1% | no | better tolerance | yes | possible | less frequent | average | yes | no | less simple | difficult | easy | difficult | Low | higher |
| Real Time Transient Model (RTTM) | seconds | yes | no | 1% | yes | best tolerance | yes | possible | less frequent | low | yes | no | most complex | more difficult | easy | more difficult | difficult | highest |
| Frequency Response | NA | yes | yes | 1% | yes | better tolerance | yes | Possible | NA | NA | yes | no | less simple | simple | easy | simple | Low | less |
| Negative Pressure Wave | seconds | no | yes | 1-5% | yes | no tolerance | no | no | frequent | NA | part time | yes | simple | easy | easy | easy | Low | less |
| Mass Balance | hours | yes | yes | 1-5% | no | no tolerance | yes | no | frequent | low | no | no | simple | yes | easy | easy | Low | less |

NA: Not Available
*some features may overlap across different classes; however, this classification is more relevant to the survey study

TABLE 3
FEATURES COMPARISON OF EXTERNAL LDS METHODS

| External Leak Detection Methods | Technical Features* | | | | | | | | | Operational Features | | | | | Economical Features | |
|---|---|---|---|---|---|---|---|---|---|---|---|---|---|---|---|---|
| | Response Time | Released Volume Estimate | Existing Leak | False Alarms | Robustness | Complexity | Affected by multiphase or multicomponent | Noise susceptibility | Leak Sensitivity | Shut-in condition | Slack condition | Availability | Ease of Use | Cont. Monitoring | Ease of retrofit | Maintenance Requirements |
| Liquid sensing | seconds to minutes | no | no | Less Frequent | High | Low | no | Low | NA | yes | yes | yes | Med | no | Difficult | Low |
| Fiber Optic Cable | seconds to minutes | yes | no | Less Frequent | Med | Low | no | Low | NA | yes | yes | yes | Med | yes | Difficult | Med |
| Vapor Sensing | minutes | yes | yes | Less Frequent | High | Low | no | Low | More Sensitive than Computational methods | yes | yes | no | Med | no | Difficult | Low |
| Acoustic Emission | Near real time | yes | no | Frequent | High | Low | yes | Med | | yes | yes | yes | Med | no | Moderate | Low |

NA: Not Available
*some features may overlap across different classes; however, this classification is more relevant to the survey study.

TABLE 4
CAPABILITIES AND LIMITATIONS OF EXISTING LDSs

| LEAK DETECTION TECHNIQUE | CAPABILITIES | LIMITATIONS | REFERENCES |
|---|---|---|---|
| Acoustic with Noise Correlation | -Popular, Easy to use<br>-Less dependent on listener skills<br>-Can work for metallic and non-metallic and large-diameter pipes<br>-Works in distribution networks | -Expensive, Labor intensive<br>-Small leaks may be missed<br>-Interference from noise<br>-Limited success in trunk mains | [20], [17], [21] |
| Acoustic-Fiber Optic | -Long Term monitoring<br>-Fiber can be additionally used for data communication purposes<br>-Excellent in detection of leaking joints and weld leaks | -Expensive<br>-Easy to break but difficult to repair | [27], [17] |
| Acoustic-Inline | -Applicable for all Pipe size and diameter<br>-Tethered accurately pinpoint leaks, non-tethered can survey long distances | -Risk of losing free swimming hydrophones<br>-Tethered hydrophone needs flow rate to flow along the pipe | [2], [30], [53], [10] |
| Broadband Electromagnetic(BEM) | -Independent of frequency so independent of electromagnetic interference<br>-Detect cracks and other anomalies | -Works only for ferrous pipes, Intensive post-processing<br>-Pipe must be drained, exposed and opened<br>-Exorbitant amount of data to be processed<br>-Manual, time consuming and labor intensive<br>-Subjective as it depends heavily on expert judgment | [10], [2] |
| Closed Circuit Television Inspection (CCTV) | -Examine pipe wall surface for defects | -Lack of consistency and reliability<br>-Surveys only above the waterline<br>-Real time assessment needed-though some new automated processing techniques | [11], [32], [8], [17] |
| Eddy Current | -Good for small metallic pipes | -Access to pipe required<br>-Skin effect limits testing only on the surface near to the probe | [52], [17] |
| Ground penetrating radars (GPR) | -Used from the surface<br>-Independent of pipe materials | -Hard to interpret, highly skilled personnel needed to interpret results<br>-Need to choose a right frequency for different soils<br>-Metal objects in ground can raise false alarms | [53], [10], [8], [12] |
| Impact Echo or Spectral Analysis of Surface waves | -Detects voids, cracks and overall condition can detect entire length of pipe investigate both pipe and soil conditions | -Thorough cleaning needed<br>-Access to pipe needed to excite the pipe Presence of tuberculation in pipe mains will render the echo dysfunctional<br>-Cannot detect extent of cracks | [8] |
| Infrared Thermography | -Used from the surface, non-invasive<br>-Can cover large areas without excavation<br>-Accurately determine geometry and defects<br>-Can scan entire length of pipe | -Useful only with liquids and gas having higher temperature than surroundings<br>-Weather restrictions such as wind speed and ground cover can influence results Expensive<br>-Significant Training and Experience needed<br>-Unable to measure depth | [17], [10], [8], [2] |
| Laser Scanning | -Reduce the cost of testing considerably<br>-Can be coupled with algorithms to classify the detects | -Inspect only dry portions of pipe<br>-Time consuming | [53], [11], [17] |
| Listening Sticks | -Simple and cheap | -Success depends on experience of user<br>-Background noise can cause erroneous detection<br>-Can only detect area of the leak, not the number and positions of leaks | [17], [8], [21] |
| Magnetic Flux Leakage (MFL) | -Exact location, size and shape of the defects<br>-Reliable, low operational costs, suitable for small diameter pipes(<=12 in)<br>-Suitable for cast iron and steel pipes | -Access to Pipe required<br>-Test results require human expertise<br>-In-line MFL has size limitation, external MFL requires costly excavation of pipes | [28], [42], [32], [8], [17], [1] |
| Remote Field Eddy Current (RFEC) | -Reliable, low operational costs<br>-Suitable for small diameter pipes(<=12 in)<br>-Only for metallic pipes | -Health and safety issues<br>-Not available for cement/asbestos pipes small leaks not detectable<br>-Pipelines need to be dewatered | [28], [1] [40] |
| RFID | -Information about utility can be embedded | -Need to attached to utilities | [17] |
| Sewer Scanning Evaluation Technology | -Post Processing of images possible<br>-Image processing and ANN available for automatic mode | -Subjective due to human interpretation of results | [8], [17] |
| Sonar or Ultrasonic | -Determines inner profile of the pipe along its length<br>-Capable of detecting pits, voids and cracks | -Only cracks perpendicular to the beam are identified, cracks parallel to beam are missed<br>-Operated in air or water but not simultaneously | [8], [42], [17], [22] |

| Method | Advantages | Disadvantages | Ref |
|---|---|---|---|
| Ultrasonic guided wave | -Quick inspection<br>-Detect corrosion<br>-Large inspection coverage<br>-Reliable, low operational costs, suitable for small diam pipes(<=12 in) | -Close contact with pipe wall needed which can damage the pipeline<br>-Complex wave pattern to be interpreted<br>-Significantly affected by environmental conditions<br>-Difficult to identify areas of cracking | [1], [17], [12] |
| Ultrasound | -Good detection rates reported for crude and oil gas pipelines for defects such as voids, cracks and corrosions<br>-Can determine location and site of defect<br>-Can detect area of leaks | -Thorough cleaning needed for inspection<br>-Cannot assure timely detection of leaks<br>-Depends on the diligence of the inspection team | [28], [8] |
| Visual Observation | -Inspection does not require any equipment or tools<br>-Capable of detecting 1-5% leaks in minutes to hours<br>-Method easy to learn and use | -Area needs to be isolated to find precise position of leaks<br>-Aerial surveys can miss small leaks<br>-Location of leaks cannot be determined<br>-False alarms during transient conditions | [17], [43] |
| Volume Balance | -Implementation on existing system or retrofitting is easy<br>-Involves less cost<br>-Estimate the volume and location of leaks<br>-Leaks detectable in shut-in conditions | -Leaks cannot be detected during shut-down or slack-in or in transient conditions<br>-Small leaks, existing leaks and leaks during slack line conditions cannot be detected | [43] |
| Rate of Pressure/Flow Change | -Able to detect 5% leak in minutes<br>-Maintenance, retrofitting is easy | -False alarms frequent during transient conditions<br>-Implementation and testing is not easy<br>-Method not easy to learn and use | [13] |
| Mass Balance with Line Pack compensation | -Existing leaks and leaks for shut in and transient conditions can be detected<br>-Able to detect 1% leak in minutes<br>-Method adaptable to any pipeline configuration | -Leaks cannot be detected during slack in conditions<br>-Implementation, maintenance and retrofitting not easy<br>-Location of leak cannot be determined<br>-Cost is high<br>-Implementation, maintenance and retrofitting is difficult | [5] |
| Real Time Transient Model(RTTM) | -Capable of detecting 1% leaks in seconds<br>-Leaks can be detected in shut-in, slack line or transient conditions<br>-Leak location and leak flow rate can be identified | -High cost<br>-Model need to be customized and tuned for each pipeline configuration<br>-Method difficult to learn and use<br>-Implementation, testing, and maintenance is difficult | [43] |
| Statistical Data Analysis | -Capable of detecting 1-5% leaks in seconds to minutes<br>-Leak location can be identified<br>-False alarms less frequent<br>-Method easy to learn and use<br>-Method easily adaptable to any pipe configuration | -Expensive technology<br>-Leaks in slack-line conditions cannot be identified<br>-Implementation and testing are difficult Costs are high | [43] |
| Fiber Optic Cables | -Fiber optic immune to electromagnetic inference, humidity, vibration and corrosion<br>-Can estimate the location of leak Response time is reasonable, responds in seconds to minutes | -Retrofitting to existing pipelines is very difficult<br>-Instability of the chemical coating possible and lead to false alarms<br>-Costs are extremely high<br>-Cable replacement may be needed after a leak occurred | [43] |
| Vapor Sensing | -Location and size of the leak can be estimated<br>-Operated in a continuous mode<br>-Responds in minutes | -Method not effective for above the ground pipes<br>-Costs are prohibitive | [43], [44] |
| Acoustic Emission | -Operated in a continuous mode and can be automated<br>-Can determine leak location and size of the leak<br>-Does not require shutdown for installation and calibration<br>-Minimally affected by computational flow<br>-Can be applied new or retrofitting to existing pipelines<br>-More sensitive than computational methods and responds in real-time | -Noise conditions such as valve noise, pump noise, multiphase flow can mask leak signal<br>Numerous sensors needed to monitor pipelines<br>Costs are high<br>-Cannot be used for already broken wires | [44], [2] |